\documentclass[sigconf]{acmart}
\AtBeginDocument{%
  \providecommand\BibTeX{{%
    \normalfont B\kern-0.5em{\scshape i\kern-0.25em b}\kern-0.8em\TeX}}}

\setcopyright{rightsretained}
\copyrightyear{2021}
\acmYear{2021}
\acmISBN{}
\acmDOI{}
\acmConference{CHI '21 Workshop: Decolonizing HCI Across Borders}{May 8--13, 2021, Yokohama, Japan}

\begin{document}

\title{Accessibility Across Borders}

\author{Garreth W. Tigwell}
\email{garreth.w.tigwell@rit.edu}
\affiliation{%
  \institution{School of Information} \institution{Rochester Institute of Technology}
  \city{Rochester}
  \state{NY}
  \country{USA}
}

\author{Kristen Shinohara}
\email{kristen.shinohara@rit.edu}
\affiliation{%
  \institution{School of Information} \institution{Rochester Institute of Technology}
  \city{Rochester}
  \state{NY}
  \country{USA}
}

\author{Laleh Nourian}
\email{ln2293@g.rit.edu}
\affiliation{%
  \institution{School of Information} \institution{Rochester Institute of Technology}
  \city{Rochester}
  \state{NY}
  \country{USA}
}

\renewcommand{\shortauthors}{Tigwell et al.}



\maketitle

\section{Introduction}
Postcolonial computing emphasizes that the design, development, and use of technology varies within different cultures, as well as the need for understanding these design practices and what implications there are in a global context~\cite{irani2010postcolonial}. Within human-computer interaction (HCI), there is also a concept called WEIRD (Western, Educated, Industrialized, Rich, and Democratic), which highlights a requirement to better understand how current HCI principles fit in other countries and it emphasizes the necessity to more critically review our processes~\cite{sturm2015how}. A CHI 2021 paper has identified that 73\% of nearly 3,300 papers from 2016 to 2020 included Western participants and, in general, fit the WEIRD profile, calling attention to the need for more international focus in HCI research studies~\cite{linxen2021how}.

\subsection{Position Statement}
Since prior work has identified that cultural differences influence user design preferences and interaction methods, as well as emphasizing the need to reflect on the appropriateness of popular HCI principles, we believe that it is equally important to apply this inquiry to digital accessibility and how accessibility fits within the design process around the world.

\subsection{Current Research Progress}
Our long-term plan is to build upon work in this area by investigating how digital designers in different parts of the world consider accessibility and whether current accessibility resources (often developed in the west) meet or conflict with their approach to design.

Acknowledging the immensity of this task, we have narrowed our focus to a smaller project spanning just a few years that will look at improving web and mobile accessibility resources for Iranian designers. This work will be led by an Iranian PhD student.

We focus on Iranian web and mobile designers for several reasons: \textbf{1)} Iran has a fairly recent history for establishing disability rights (2009) vs the US (1973); \textbf{2)} There are distinct differences between Farsi and English (e.g., reading/writing order and grammatical structure), which likely influence design layout variances and could affect accessibility; \textbf{3)} There is scarce prior work that compares accessibility and usability practices of Iranian designers with reported US practices; and \textbf{4)} WCAG---the gold standard web and mobile accessibility resource---is not available in Farsi and a straightforward translation may not be a sufficient solution.

To understand the current accessibility support available to Iranian web and mobile designers, we plan to follow a contextual design approach using established qualitative methods and contextual inquiry~\cite{holtzblatt2014contextual}. We can then begin to identify how best to implement new accessibility guidelines and design tool features that best meet the needs of Iranian designers. However, due to COVID-19, our studies will be conducted remotely (e.g., online questionnaires, online semi-structured interviews, remote diary studies).
\section{Background and Motivation}
\subsection{Digital Accessibility}
Websites and mobile apps are important in society and enhance how we communicate, learn, socialize, and work~\cite{bjorn2014does,bohmer2011falling,hill2001teaching}; it is critical to make digital spaces accessible for people with disabilities who make up just over 1 billion people worldwide~\cite{WHOFigures}.

Although designers can make digital spaces accessible through a combination of following guidelines, using appropriate design tools, and conducting user evaluations~\cite{lazar2015ensuring}, there is a significant body of work that continues to identify the persistence of inaccessibility for both web and mobile technology (e.g.,~\cite{hanson2013progress,kuzma2010accessibility,patra2014quantitative,ross2018examining}). Web accessibility-related lawsuits are also increasing~\cite{usablenet2018}, indicating that more research is required to improve a worsening situation by better supporting web and mobile designers. Some reasons for inaccessibility include: a lack of sufficient education on accessibility in design, limited project time and funds, clients who prioritize other design requests over accessibility, confusing guidelines, and inadequate tools (see:~\cite{crabb2019developing,li2021accessiblity,patel2020why,swallow2014speaking,tigwell2017ace,tigwell2018designing,tigwell2021nuance}). Overall, inaccessibility is never usually the result of a single issue but one that can be traced back to several factors falling on a spectrum of intrinsic to extrinsic causes~\cite{ross2017epidemiology}, such as a developer’s software not warning about accessibility violations (intrinsic) and a company adhering to a culture where the importance of product accessibility is dismissed as being unimportant or too costly (extrinsic). It is necessary to explore all possible causes for the widespread issue of inaccessibility in digital spaces and then develop solutions to address those barriers. 

One under-researched area is a critical evaluation of the suitability of established accessibility practices for international designers who have different cultural norms, language systems, perspectives, and understandings. Prior work has demonstrated that cultural differences can influence design and design preferences (e.g.,~\cite{reinecke2014quantifying}), while design can be closely linked to the accessibility of the system because it is a significant part of the interface we interact with (e.g., colors, font style, layout, navigation structure)~\cite{petrie2004tension}. Identifying international differences and challenges is an important first step toward the introduction of protocols that could result in a more synergistic global attitude toward making digital content accessible.

\subsection{Cultural and Visual Design}\label{culture-visual-design}
The aesthetic design of a digital space is often intentionally created in a way to complement its purpose and evoke specific emotions~\cite{ling2002effect,cyr2010colour,bonnardel2011impact}. Yet, people from different cultures can vary in their preference for use of color and amount of text~\cite{evers1997the}, how information is organized~\cite{callaha2005cultural}, and the design of icons~\cite{kim2005cultural}. Design must include both elements that are considered universal to improve usability on a global scale and specific localized design elements~\cite{gu2016east}. Websites can be designed to represent the cultural values of the country the designer is from and it can result in distinct design differences~\cite{singh2005analyzing}. This should motivate designers to carefully consider how content is created to appeal to different audiences~\cite{khaslavsky1998integrating,russo1993how}. 

The effect of culture runs deeper than simply visual design preferences and it can influence how people seek information, navigation behaviors, and decision-making outcomes when people work in a group~\cite{el1997technology,kralisch2004cultural,kralisch2005impact,reinecke2013doodle}. Visual and interaction design preferences may conflict with accessibility recommendations.

\subsection{Cultural Differences and Usability Testing}\label{culture-usability-testing} Chavan~\cite{chavan2005another} argues that little work has been done to adapt design process methods and tools for cultural differences. Common design methodologies used in industry include user-centered design \cite{lowdermilk2013} and co-design~\cite{prahalad2004co}, where designers consider and involve future users in the design process, and running evaluations can help with design refinement~\cite{tan2009web}. User involvement may result in misunderstandings or socio-cultural conflicts without proper planning. 

For example, a series of studies conducted in the Netherlands and South Korea demonstrated that different factors and expectations need to be adapted to maintain similar levels of participant engagement~\cite{van2006three}. Another study on design method tasks also found differences in levels of engagement and forthrightness (Korean participants were less spontaneous) due to cultural norms, but this could be overcome when applying appropriate adjustments (e.g., by increasing communication to boost motivation)~\cite{lee2007cultural}.

Evaluations may also use standardized instruments to record data to inform the development of a design. The System Usability Scale (SUS) is a popular evaluation metric created in English and used within HCI since the 1980s, which takes the form of a standardized questionnaire and allows participants to indicate their feelings toward the usability of a system~\cite{lewis2018the}. There have been translations of this tool so that users who are not native speakers can use the test. Dianat et al.~\cite{dianat2014psychometric} created a translated version of the SUS into Farsi for Persian participants. In addition to the translation, the Persian SUS was evaluated with 202 participants and 10 experts and determined it remained reliable~\cite{dianat2014psychometric}. This work demonstrates the need to adapt standardized procedures and also that it can be achieved successfully, thus resulting in better-designed systems because they can be evaluated more effectively.

\subsection{Accessibility and Cultural Differences}
Sections~\ref{culture-visual-design} and~\ref{culture-usability-testing} cover many design-related cultural differences studies that focus on usability. We want to build upon this and investigate the effects on accessibility, which is unknown since it was not the focus of prior work. 

Usability is often determined by a person’s expertise and prior experiences, whereas accessibility depends on whether a user can complete tasks regardless of their abilities~\cite{keates2003countering}. To expand on this distinction with some examples, a system that is designed to be more usable will cater to not only expert users, but also novice users, thus making it easy to complete tasks (e.g., designers add clear labeling to buttons and menus to support new users). A system that is designed to be accessible provides alternative access and avoids creating barriers for users (e.g., designers avoid assigning meaning to colors and will use high contrasting colors to improve color distinction for color blind users). However, these two dimensions can be in conflict. For example, if a designer uses colors with culturally assigned meaning to improve usability (e.g., green = good, red = bad), then it will be inaccessible for people with red-green color blindness. Similarly, research in the previous section discussed cultural preferences for layout and amount of text, which could conflict with Western-defined accessibility guidelines.

Since 1999, WCAG~\cite{kirkpatrick2018web} has become the gold standard for ensuring the accessibility of websites and mobile apps (e.g., iOS~\cite{iOS}), but it has not always been clearly implemented within law and policy, thus challenging the implementation of accessible digital spaces~\cite{lazar2019web}. A simple translation of WCAG may also be ineffective, since, as an accessibility resource, WCAG is already often criticized by designers for being difficult to use~\cite{swallow2014speaking}. Instead, there may be an opportunity to work closely with international designers to establish a more culturally sensitive set of accessibility guidelines.

A collection of research has investigated the accessibility of different government websites -- examples include India~\cite{patra2014quantitative}, Pakistan~\cite{bakhsh2012web}, Saudi Arabia and Oman~\cite{abanumy2005government}, South America~\cite{lujan2014egovernment}, Taiwan~\cite{huang2003usability}, UK~\cite{kuzma2010accessibility}. The research suggests that reasons for these accessibility violations may be caused by a lack of laws and policies~\cite{bakhsh2012web,lujan2014egovernment}, but it is unknown whether those designers are adequately supported in meeting accessibility guidelines because the work primarily took a quantitative methods approach. We can leverage the advantages of qualitative methods to better understand the needs and concerns of international designers.
\section{Our workshop goals}
We have several reasons for wanting to attend the CHI 2021 Workshop: Decolonizing HCI Across Borders. We would like to:

\begin{itemize}
    \item Get feedback on our project to refine the research plan.
    \item Receive guidance on how to navigate through this work in a culturally sensitive way. Although our PhD student is Iranian, we will likely look at different cultures so that we can understand similarities and differences.
    \item Understand how best to conduct remote work across borders (e.g., \textit{what are the common challenges to consider?} and \textit{how do researchers in this area overcome those challenges?}).
    \item Learn about the most up-to-date work that is being led by experts in this research area.
    \item Network with other researchers who are passionate about `decolonial' thinking within HCI.
\end{itemize}

In addition to our own needs, we also want to contribute to the success of the workshop through active participation and offer feedback on other work that is presented. We can share with the attendees our knowledge on web and mobile design, digital designer's work practices, and accessibility, in addition to our expertise in qualitative data collection methods.

\bibliographystyle{ACM-Reference-Format}

\end{document}